\documentclass[utf8]{jetp}
\usepackage{lmodern}
\usepackage{amssymb}
\usepackage{amsmath}
\allowdisplaybreaks
\usepackage{graphicx}
\usepackage{bm}

\twocolumn
\UnFig
\graphicspath{{./figures/}}

\renewcommand\vec{\bm} 
\newcommand{\uGrad}{\vec{\nabla}}

\newcommand{\ud}{\,{\mathrm{d}}}

\newcommand{\uiiint}{\int\!\!\!\int\!\!\!\int}

\newcommand{\uC}{C} 
\newcommand{\uD}{D} 
\newcommand{\ue}{e} 
\newcommand{\uer}{e_{\rho}}
\newcommand{\uet}{e_{\theta}}
\newcommand{\uE}{E} 
\newcommand{\uF}{F} 
\newcommand{\uFc}{\cal F} %

\newcommand{\uHopf}{{\cal{H}}} 
\newcommand{\uh}{h} 
\newcommand{\uH}{H} 
\newcommand{\uvH}{\vec{H}}     
\newcommand{\uK}{K}         

\newcommand{\uvM}{\vec{M}}
\newcommand{\uvm}{\vec{m}}
\newcommand{\umi}{m_i}      
\newcommand{\umx}{m_{\mathrm{X}}}
\newcommand{\umy}{m_{\mathrm{Y}}}
\newcommand{\umz}{m_{\mathrm{Z}}}

\newcommand{\uO}{O}         

\newcommand{\upex}{p_\mathrm{EX}}
\newcommand{\upa}{p_\mathrm{A}}
\newcommand{\upz}{p_\mathrm{Z}}
\newcommand{\updm}{p_\mathrm{DM}}

\newcommand{\uMs}{M_{\mathrm{S}}}

\newcommand{\umuZ}{\mu_0}
\newcommand{\uq}{q} 
\newcommand{\uvs}{\vec{s}}  
\newcommand{\uX}{X}         
\newcommand{\uY}{Y}         
\newcommand{\uZ}{Z}         
\newcommand{\uXp}{\widetilde{X}} 
\newcommand{\uYp}{\widetilde{Y}} 
\newcommand{\uZp}{\widetilde{Z}} 
\newcommand{\uR}{R}    
\newcommand{\ur}{r}
\newcommand{\urp}{\widetilde{r}}
\newcommand{\uvr}{\vec{r}}          
\newcommand{\uvrp}{\widetilde{\vec{r}}}  
\newcommand{\ut}{\theta}

\begin{document}
\English
\def\TitleSkip{\vskip 5mm}
\def\AuthorsSkip{\bigskip}
\def\AffilSkip{\medskip}
\def\AffiliationsSkip{\medskip}
\def\DatesSkip{\bigskip}
\def\DOISkip{\vskip -5mm}
\def\LFVS{\vskip -5mm}
\def\ELFVS{\vspace*{0mm}}

\title{Control of the magnetic hopfion lattice in helimagnet with the external field and anisotropy}

\rtitle{Control of the magnetic hopfion lattice\dots}

\affiliation{Galkin Donetsk Institute for Physics and Engineering, R.~Luxembourg
 str.~72, Donetsk 283048, Russian Federation}

\author{K.~L.}{Metlov}\email{metlov@donfti.ru}
\author{A.~S.}{Tarasenko}
\author{Yu.~A.}{Bezus}
\author{M.~M.}{Gordei}

\dates{29 Match 2025}{*}
\abstract{A generalized micromagnetic model of hopfions in a helimagnet with a two-dimensional (allowing both radial and azimuthal dependence) profile function is considered. Calculations confirm the elliptical stability of hopfions and the previously obtained analytical expression for the upper critical field of their lattice. Dependencies of the hopfion lattice periods on the magnitude of the applied external magnetic field and the uniaxial anisotropy constant of the material are obtained. It is shown that in an anisotropic helimagnet, the hopfion lattice expands in the direction of the anisotropy axis, and the expansion can be controlled by the external field and the uniaxial anisotropy constant.}

\DOI 10.7868/XXXXXXXXXXXXXXX

\maketitle
	
\section*{Introduction}
Magnetic topological solitons~\cite{KIK83-book} play a key role in almost all applications of magnetism from transformer cores, sensors, and hard disk platters to solid-state magnetic memory and nanoscale microwave generators. From the topological point of view, they all correspond to continuous mappings of a sphere onto a sphere. The target sphere is always $S^2$, on which the endpoints of the fixed length local magnetization vectors $|\uvM(\uvr)|=\uMs$ are located. The source sphere is a mapping of the physical space $\uvr$ with periodic boundary conditions imposed at infinity. It is known that in one ($S^1\rightarrow S^2$) and two ($S^2\rightarrow S^2$, \cite{BP75}) dimensions such mappings split into integer-numbered homotopy classes. Thus, discrete localized objects (domain walls~\cite{Hubert_book_walls}, cylindrical magnetic domains~\cite{BG88}, skyrmions~\cite{BY1989}, magnetic vortices~\cite{UP93,M10}) arise in the magnetization vector field, and the class number plays the role of the count of these topological solitons in the system.

For a long time in mathematics it was believed that all mappings $S^3\rightarrow S^2$ are homotopically equivalent (which would mean that three-dimensional topological solitons cannot exist). However, in 1931 Heinz Hopf gave an example of a mapping~\cite{Hopf1931} that is not reducible to all the others. This example was generalized by Whitehead~\cite{whitehead1947}, who showed that in this case too the mappings split into homotopy classes numbered by integers, introduced the corresponding Hopf index $\uHopf$ (the original counter-example~\cite{Hopf1931} has $\uHopf=1$) and constructed mappings with arbitrary $\uHopf$. The idea of searching for solitons, corresponding to these mappings, in physics was expressed by Ludwig Faddeev~\cite{Faddeev1976}, and specifically in the vector fields of local magnetization by Dzyaloshinsky and Ivanov~\cite{DI79}. But this idea did not take off at the time due to the Hobart-Derrick theorem~\cite{Hobart1963,Derrick1964}, which proved instability of three-dimensional solitons in a classical ferromagnet.

In recent years, there has been a renewed interest to 3D topological solitons~\cite{RKBDMB2022} in chiral magnets (outside the applicability~\cite{BK11-book} of the Hobart-Derrick theorem). This is due to the development of methods for observing magnetization distributions at the nanoscale, such as high-resolution X-ray microscopy~\cite{KRRCVDCHSPFHSF2021}, vector magnetic nanotomography~\cite{CMSGHBRHCG2020} and magnetic small-angle neutron scattering~\cite{Metlov2024}, for which magnetic hopfions would be an ideal object of study. Another incentive is the potential for hopfions to be used in 3D spintronics~\cite{BK11-book,RKBDMB2022,Gubbiotti2024roadmap}.

The first experimental observations of hopfions and hopfion-like states in confined geometry have appeared~\cite{KRRCVDCHSPFHSF2021,zheng2023,Yu2023}. They are also reproduced qualitatively in numerical experiments~\cite{lake2018,TS2018,sutcliffe2018,WQB2019,RKBDMB2022}. However, these are not free-standing bulk hopfions, but rather depend on external stabilization and confinement for their existence. This is typically achieved by sandwiching a hopfion-containing film between two magnetic layers with strong perpendicular magnetic anisotropy~\cite{TS2018}. However, the true potential of hopfions can only be realized in a fully three-dimensional "bulk" environment.

Our starting point here is a variational model of bulk hopfions in a helimagnet~\cite{M2023_TwoTypes} based on the standard micromagnetic Hamiltonian taking into account the Dzyaloshinskii-Moriya interaction. It predicts the existence of two (topologically equivalent) types of magnetic hopfions. Only one of them is stable with respect to elliptical deformation~\cite{M2025} and can exist unconfined in a bulk magnet, the other is not. Hopfions of the second type tend to expand indefinitely along the direction of the external magnetic field, maintaining a stable size in the perpendicular direction. This expansion can be stopped by introducing layers with increased uniaxial magnetic anisotropy into the system~\cite{TS2018}. Hopfions of the first type do not require such artificial stabilization.

In the present paper, the model~\cite{M2025} is generalized by introducing an additional dimension (corresponding to an additional infinite set of variational parameters) into the trial function. Such an extended model reduces to the original one in a limiting case, but by definition (due to a larger number of free parameters) gives a better approximation to the energy of the exact solution. A numerical estimate of the upper critical field of the Hopfion lattice is given, which turns out to be in an excellent (taking into account the numerical error) agreement with the previously obtained analytical expression~\cite{M2025}. The considered extended model also allows one to ``untie'' the internal structure of the hopfion from its external contour, and therefore opens up the possibility of modeling the contact (exchange, since the Dzyaloshinskii-Moriya interaction is also a part of it) repulsion between hopfions in the lattice. This makes it possible to study the influence of the external field and anisotropy on the parameters (periods) of the equilibrium hopfion lattice.

\section{Model}
\label{sec:model}
Consider a classical helimagnet with free energy per unit volume $\uE=\lim_{V\rightarrow\infty} (1/V) \uiiint_V F \ud^3\uvr$ with density
\begin{align}
\uF = &\frac{\uC}{2}\sum_{i=\uX,\uY,\uZ} \left|\uGrad\umi\right|^2 +\uD\, \uvm\cdot[\uGrad\times\uvm] - \nonumber \\
&\umuZ \uMs \left(\uvm \cdot \uvH\right) - \uK\left(\uvm \cdot \uvs\right)^2,
\label{eq:energy}
\end{align}
where ${\bf m}({\bf r})={\bf M}({\bf r})/\uMs$, $|\uvm|=1$, $\uC=2A$ is the exchange stiffness in J/m, $\uD $ is the Dzyaloshinskii-Moriya (DM) interaction parameter~\cite{BS1970,BJ1980} in $\text{J}/\text{m}^2$,\ $\uK$ is the uniaxial\ (directed along $\uvs$) anisotropy constant in $\text{J}/\text{m}^3 $, $\uvH$ is the external magnetic field in A/m, and $\umuZ=4\pi\times{10^{-7}}$ H/m is the vacuum permeability. For simplicity, we assume that the field and the anisotropy axis are parallel, and choose a Cartesian coordinate system such that this direction coincides with the $\uO\uZ$ axis: $\uvH=\{0,0,H\}$, $\uvs=\{0,0,1\}$. The problem now is to find a vector field $\uvm(\uvr)$ that minimizes $\uE$ and has $\uHopf>0$.

The exact solution to this problem is not known today. Therefore, we will use the Ritz method, choosing as a starting point the following trial function for the magnetization vector field:
\begin{subequations}
\label{eq:model}
\begin{align}
\label{eq:stereogr}
&\{\umx+\imath\umy,\umz\}=\{2w, 1 - |w|^2\}/(1+|w|^2),\\
\label{eq:whitehead}
&w=\imath u/v,\\
\label{eq:E3S3}
 &u=\frac{2(\uXp+\imath \uYp)\uR}{\urp^2+\uR^2},
 \ 
 v=\frac{\uR^2-\urp^2+\imath 2 \uZp \uR}{\urp^2+\uR^2},
\\
\label{eq:physE3}
&\uvrp=\uR\,\frac{\uvr}{r}\frac{\ue}{1-\ue},\ \ue=\ue\left(\frac{r}{\uR},\arctan\frac{\uZ}{\sqrt{r^2-\uZ^2}}\right)
\end{align}
\end{subequations}
$\uvr=\{\uX,\uY,\gamma\uZ\}$, $r=|\uvr|$, $\uvrp=\{\uXp,\uYp,\uZp\}$, $\urp=|\uvrp|$. It defines $\uvm(\uvr)$ with Hopf index $\uHopf=1$, parameterized by a (still unknown) profile function $\ue(\rho,\theta)$, $\rho=\ur/\uR$, satisfying the boundary conditions $\ue(0,\theta)=0$, $\ue(1,\theta)=1$, and two scalar parameters -- the Hopfion radius $\uR$ and the shape factor (the ratio of the size in the direction perpendicular to the magnetic field to the size in the direction of the field) $\gamma$. The hopfion is completely contained within the ellipsoid $\rho<1$, and on its boundary ($\rho=1$) has a magnetization $\uvm=\{0,0,1\}$, which coincides with the magnetization outside the hopfion. When $|\gamma|<1$, the ellipsoid is elongated along the $\uO\uZ$ axis.

The central to this trial function is the Whitehead's ansatz, which is written here in the special case $\uHopf=1$. In general, this ansatz has the form~\cite{whitehead1947} $w=\imath u^m/v^n$, where $n$ and $m$ are relatively prime integers, which corresponds to magnetization distributions with Hopf index $\uHopf=m n$. Pre-factor $\imath$ in~\eqref{eq:whitehead} selects the first of two types of hopfions~\cite{M2023_TwoTypes} (the second type of hopfions corresponds to the factor $-\imath$ and requires additional stabilization to compensate for their elliptical instability~\cite{M2025}). The hopfions $\uHopf=1$ are axially symmetric with respect to $\uO\uZ$ and consist of circular vortex and antivortex filaments wrapped around $\uO\uZ$ with opposite magnetization at the centers of their cores. In the hopfions of the first type considered here, the outer antivortex filament is wrapped around the inner vortex filament, as shown in
\begin{figure}[h]\vspace{1mm}
\begin{center}
\hspace{0mm}
\includegraphics[width=\columnwidth]{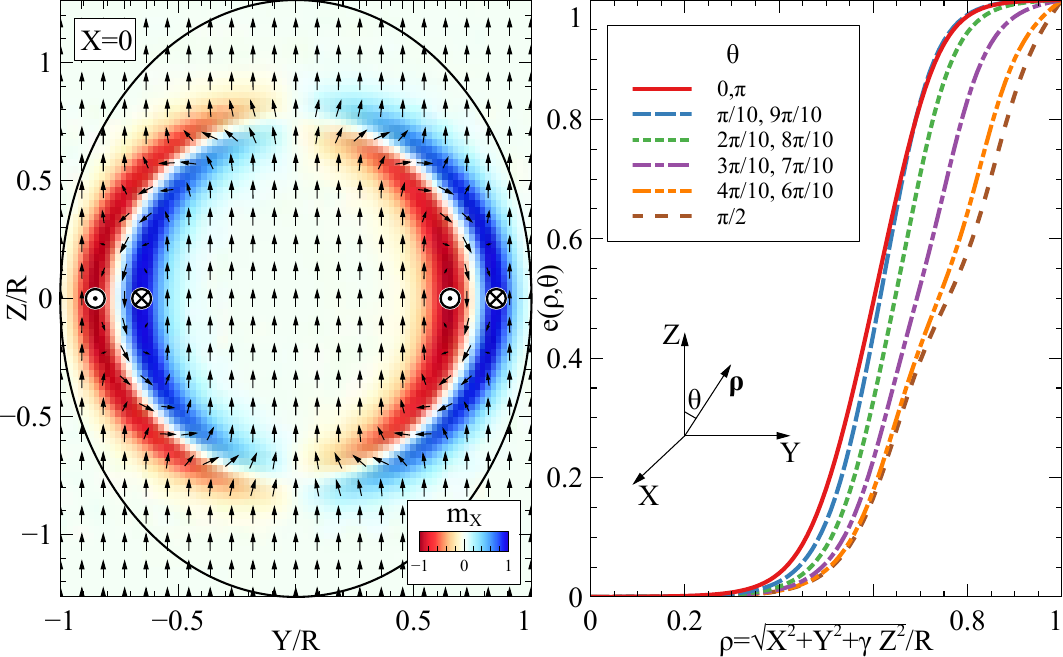}
\end{center}
\caption{\hspace{0mm} Cross-section of magnetization distribution (left) and some slices of profile function $e(x,\theta)$ (right) of equilibrium Hopfion (its overall chirality corresponds to the material with $\uD<0$) with $\uHopf=1$ at $q=1$ and $h=0.1$.}
	\label{fig:H1}
	\vspace{5mm}
\end{figure}	
Fig.~\ref{fig:H1} on the left. For hopfions of the second type, the order of the filaments is reversed. The Whitehead's ansatz~\eqref{eq:whitehead} defines a map of the sphere $S^3$ parameterized by two complex coordinates $u$ and $v$ (such that $|u|^2+|v|^2=1$) onto the Riemann sphere $S^2$ parameterized by a complex coordinate $w$. The formulas~\eqref{eq:E3S3} express $u$ and $v$ in terms of coordinates in the extended Euclidean space $\{\uXp,\uYp,\uZp\}$, and the stereographic projection~\eqref{eq:stereogr} relates $w$ to the components of the magnetization vector of fixed length $|\uvm|=1$. Together, equations~\eqref{eq:stereogr},~\eqref{eq:whitehead},~\eqref{eq:E3S3} define a vector field $\uvm(\uXp,\uYp,\uZp)$ containing one $\uHopf=1$ hopfion filling the entire space. These formulas are universal in the sense that any $\uHopf=1$ hopfions (not only magnetic ones) are necessarily reduced to them by some continuous mapping (homotopy).

The main approximations of the considered model are contained in the equation~\eqref{eq:physE3}, which defines the mapping of the physical space $\{\uX,\uY,\uZ\}$ onto the extended Euclidean space $\{\uXp,\uYp,\uZp\}$. It completely localizes the hopfion inside the ellipsoid $\rho<1$, mapping (due to the boundary condition $\ue(1,\ut)=1$) points on its boundary $\rho=1$ to an infinitely distant point in space $\uvrp$, where the magnetization vector $\uvm=\{0,0,1\}$ is directed parallel to the $\uO\uZ$ axis. The appearance of $\ue$ in the numerator of~\eqref{eq:physE3} improves the original variational model~\cite{M2023_TwoTypes} and is discussed in~\cite{M2025}. The key novelty of the variational model in the present paper is the inclusion of a possible azimuthal dependence of the Hopfion profile $\ue(\rho,\ut)$, while in the earlier studies the profile was assumed to be independent of $\ut$.

Now suppose that such localized hopfions are arranged in an infinite three-dimensional close-packed lattice (FCC or HCP) with volume $V=4\sqrt{2}R^3\gamma$ per hopfion. Given that each individual Hopfion has $\uHopf=1$, the Hopf index of the lattice as a whole is equal to infinity. The total energy of a unit volume, normalized to $D^2/C$, can be represented as $\uE\uC/\uD^2=\int_0^1\int_{0}^{\pi}{\cal F}\ud\ut\ud \rho$, where
\begin{equation}
 \label{eq:energyF}
 {\cal F}= \frac{1}{\kappa} \left[\nu^2\upex+\nu\,\updm+\frac{q}{2}(\upa\!-\!\kappa)+h(\upz\!-\!\kappa)\right],
\end{equation}
$\kappa=4\sqrt{2}$, $\nu=\uC/(\uD \uR)$ is a dimensionless parameter inversely proportional to the hopfion size $\uR$, $\uh=\umuZ\uMs\uC\uH/\uD^2$ is the normalized external field, $\uq=2\uC\uK/\uD^2$ is the normalized quality factor of uniaxial anisotropy. Expressions for the functions $\upex$, $\updm$, $\upa$, and $\upz$ are given in the supplementary section~\ref{sec:functions}. The energy $\uE$ is a functional of the hopfion profile $\ue(\rho,\ut)$ and depends on two additional dimensionless scalar parameters --- $\nu$ and $\gamma$.

Minimization of energy $\uE$ in this work was carried out numerically in two stages. At the first stage, the equilibrium profile of the hopfion $\ue(\rho,\ut)$ was calculated by the finite element method for given $\uq$, $\uh$, $\nu$ and $\gamma$ as a solution to the Euler-Lagrange equation:
\begin{equation}
 \label{eq:EulerLagrange}
 \frac{\partial \uFc}{\partial \ue} -
 \frac{\partial}{\partial\rho}
 \frac{\partial \uFc}{\partial\uer} -
 \frac{\partial}{\partial\theta}
 \frac{\partial \uFc}{\partial\uet} = 0,
 \end{equation}
 where $\uer=\partial \ue(\rho,\ut)/\partial\rho$, $\uet=\partial \ue(\rho,\ut)/\partial\ut$ with boundary conditions $\ue(0,\ut)=0$, $\ue(1,\ut)=1$. For this purpose, the {\tt NDSolve`FEM`} function of the Mathematica\texttrademark{} computer algebra system was used. To stabilize the numerical calculation, an additional diffusion term $0.0025(\partial^2\ue/\partial\rho^2+\partial^2\ue/\partial\theta^2)$ was introduced into the Euler-Lagrange equation, which has virtually no effect on the value of the hopfion energy, but suppresses high-frequency numerical fluctuations of the profile around $e\approx0$. Then, for the resulting profile, the normalized energy $\uE\uC/\uD^2$ was calculated by direct numerical integration~\eqref{eq:energyF} and minimized numerically using the function {\tt FindMinimum} over the remaining two scalar parameters $\nu$ and $\gamma$. An example of a hopfion calculated in this way and the corresponding profile functions (for chosen values of $\ut$) are shown in Fig.~\ref{fig:H1}.

 \section{Results}
 The main feature of the considered model is very well visible already in Fig.~\ref{fig:H1}: the core of the hopfion (the region of highly inhomogeneous magnetization) is no longer tied to its outer contour $\rho=1$ (shown by the black ellipse in Fig.~\ref{fig:H1} on the left). Above and below the core, regions of almost uniform magnetization are formed, but in such a way that the core of the hopfion itself (and this effect is preserved for other $\uq$ and $\uh$) retains an almost ideal spherical shape (while the outer contour of the hopfion acquires a significant elliptical deformation). This is in agreement with the conclusions obtained within the framework of a simpler model~\cite{M2025}, and confirms the elliptical stability of hopfions.

In the region $\uq>0$, $\uh>0$, when the hopfions are most energetically favorable, their stability region is limited by the upper critical field (which becomes $0$ at some finite value of the anisotropy quality factor $\uq$). In weak (subcritical) external fields, the energy of the equilibrium hopfion lattice is lower than the energy of the state with uniform magnetization. At the critical line, these energies become equal and a mutual (hysteresis-free) transformation of the hopfion lattice into the state of uniform magnetization occurs via a second-order phase transition.

When approaching the critical line, the energy minimum for the hopfion radius turns into a saddle point and disappears, and the radius itself tends to infinity (the uniformly magnetized hopfion inner core expands and fills the entire space). This means that the numerical procedure for finding the minimum of the total hopfion energy in $\nu$ and $\gamma$ stops working near the critical line. However, it is possible to get close enough to it to use linear (in the field $\uh$) extrapolation of energy to the point where the hopfion energy becomes exactly equal to the energy of the uniformly magnetized state. The results of this extrapolation are shown as dots at
\begin{figure}[tb]\vspace{1mm}
\begin{center}
\hspace{0mm}
\includegraphics[width=\columnwidth]{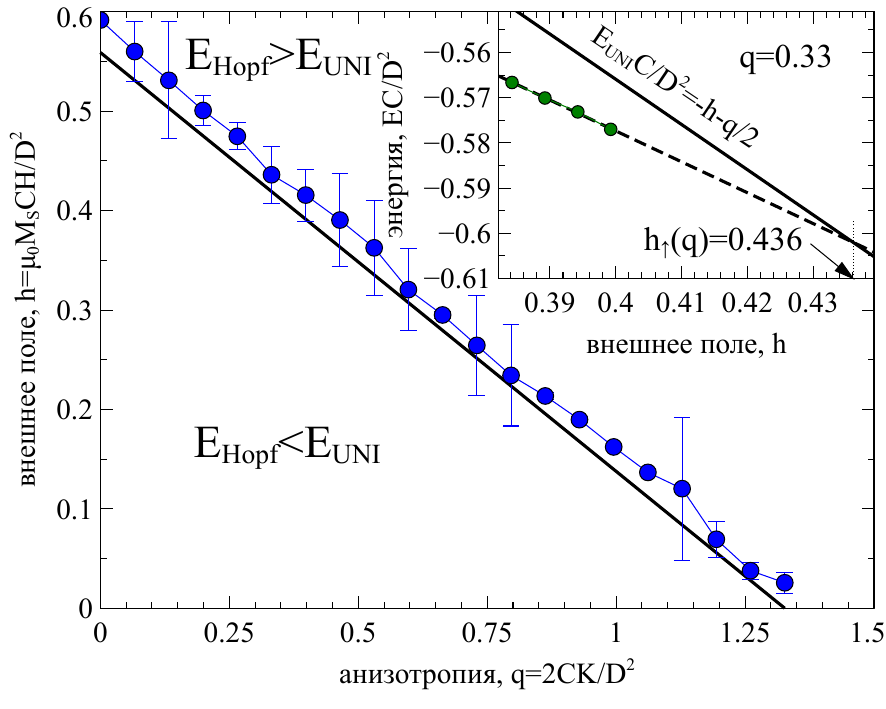}
\end{center}
\caption{\hspace{0mm} The upper critical field of the Hopfion lattice, corresponding to its transition to a state with uniform magnetization; the inset illustrates the extrapolation procedure using the example of one point with $q=0.33$.}
	\label{fig:hcrupper}
	\vspace{5mm}
\end{figure}	
Fig.~\ref{fig:hcrupper}. The errors shown take into account only the statistical spread of the four points used in extrapolation for each value of $\uq$.
The solid line in Fig.~\ref{fig:hcrupper} shows the critical field calculated using the formula
\begin{equation}
\label{eq:hup}
 h_{\uparrow}=\frac{4 (2+\pi ) (3 \pi -4) q-15 \pi ^2}{60 (\pi -4) (2+\pi )},
\end{equation}
obtained in~\cite{M2025}. The slight systematic numerical overestimate of the critical field is apparently due to the additional diffusion term introduced into the equation. But in general, it can be said that the numerical estimate of the critical field is in very good agreement with the analytical formula.
Considering that such agreement is present for three different hopfion models:~\cite{M2023_TwoTypes}, \cite{M2025} and the current one -- it can be said that the estimate~\eqref{eq:hup} if not an exact result, is, probably, very close to it.

The numerical procedure, described in section~\ref{sec:model}, allows one to calculate the equilibrium values of the parameters $\nu$ and $\gamma$ in the entire stability region of hopfions shown in Fig.~\ref{fig:hcrupper}. From these, one can estimate the influence of the external field and anisotropy on the radius (in the $\uZ=0$ plane) $\propto1/\nu$ and the height (size in the field direction) $\propto1/(\gamma\nu)$ of the hopfion. These dependencies are shown in
\begin{figure}[tb]\vspace{1mm}
\begin{center}
\hspace{0mm}
\includegraphics[width=\columnwidth]{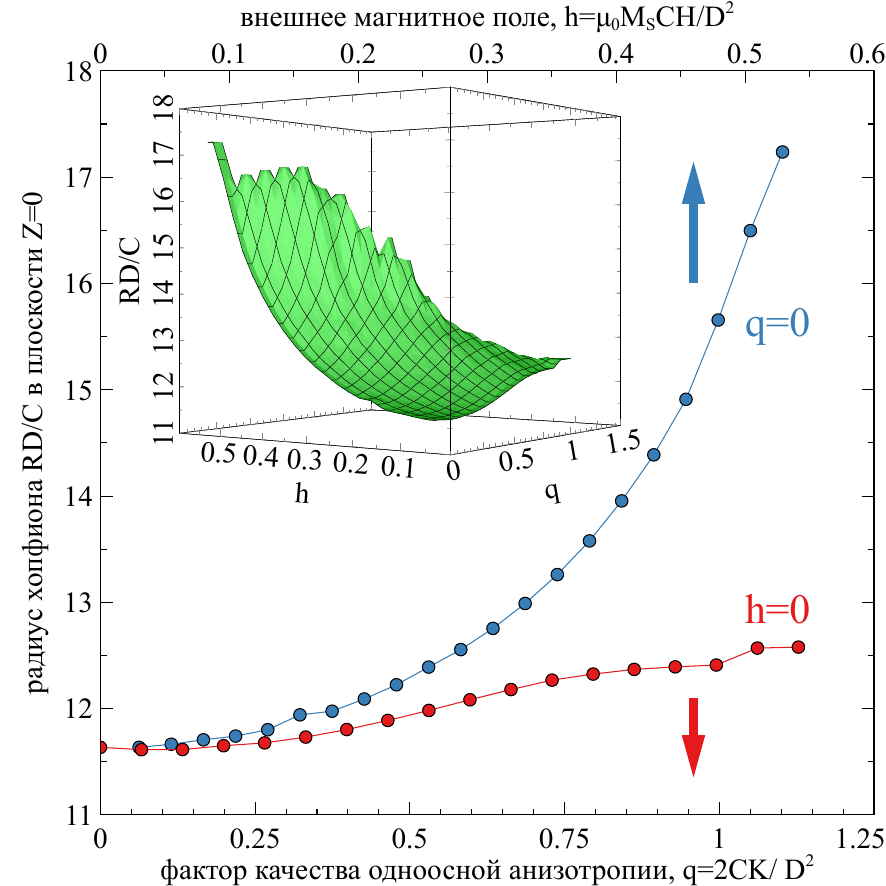}
\end{center}
\caption{\hspace{0mm}Hopfion radius as a function of the applied magnetic field and the uniaxial anisotropy quality factor at $q=0$ and $h=0$; the inset shows the full plot over the entire stability region of hopfions at $h>0$, $q>0$.}
	\label{fig:width}
	\vspace{2mm}
\end{figure}	
\begin{figure}[tb]\vspace{0mm}
\begin{center}
\hspace{0mm}
\includegraphics[width=\columnwidth]{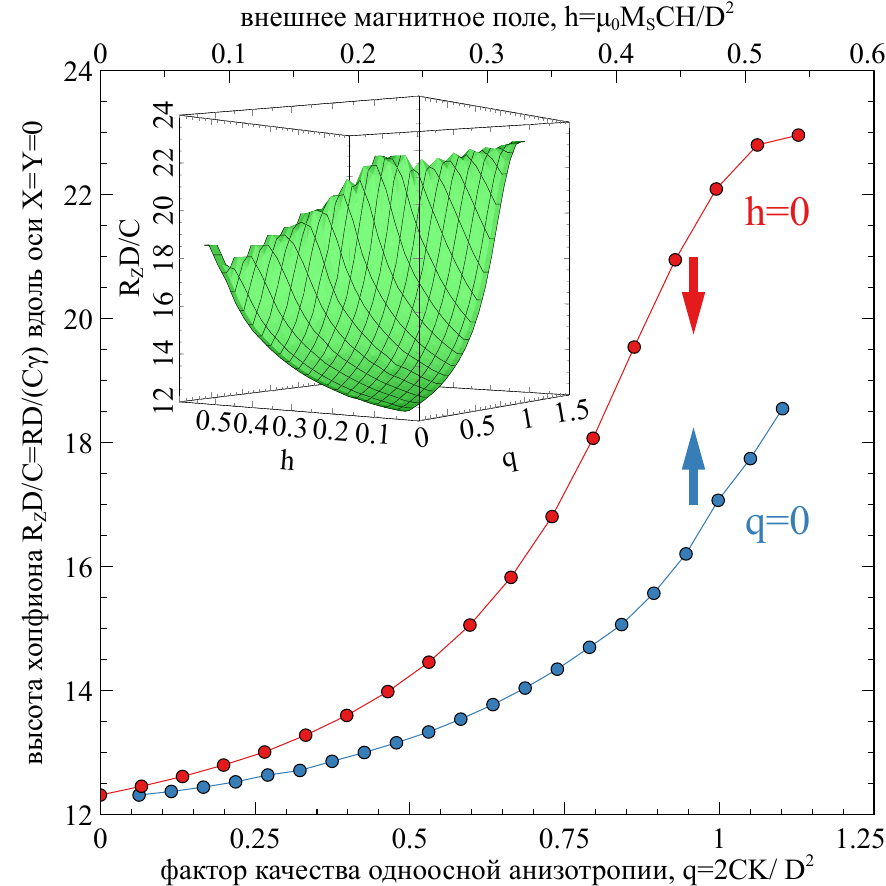}
\end{center}
\caption{\hspace{0mm}Hopfion height as a function of applied magnetic field and uniaxial anisotropy quality factor; the plot composition is identical to Fig.~\ref{fig:width}.}
	\label{fig:height}
	\vspace{4mm}
\end{figure}	
Fig.~\ref{fig:width} and Fig.~\ref{fig:height}. It is seen (especially clearly on the graphs at $\uq=0$) that a change in the external field causes approximately the same change in the radius and height of the hopfion, and a change in the anisotropy quality factor has practically no effect on the hopfion radius (see Fig.~\ref{fig:width} at $\uh=0$), but greatly affects its height (see Fig.~\ref{fig:height} at $\uh=0$). The hopfion itself (the region where the magnetization deviates significantly from uniform) retains an almost ideal spherical shape, only its outer contour changes (shown as an ellipse in Fig.~\ref{fig:H1}). However, even small deviations of magnetization inside the hopfion contour from the ideal vertical ($\uvm=\{0,0,1\}$ outside the hopfion) produce repulsive exchange forces.

Since the energy of a stable hopfion is lower than the energy of a uniformly magnetized state, the lattice of hopfions will be closely packed to minimize the space (with uniform magnetization) between them. The problem of close packing of ellipsoids has a number of non-trivial solutions~\cite{Donev2004} that do not reduce to a rescaled lattice of close-packed spheres. However, in all these unusual cases, the ellipsoids are not coaxial. Hopfions in the absence of anisotropy have an almost ideal spherical shape, and acquire a noticeable elliptical distortion only in an anisotropic magnet. But this same anisotropy imposes a preferred direction on them, and therefore prevents the realization of more complex close-packed states~\cite{Donev2004}.

In general, this means that the external field uniformly scales the lattice of close-packed spheres (and the sizes of each hopfion separately), and the presence of uniaxial anisotropy scales it more strongly in one direction. Such a stretching of the hopfion lattice in an anisotropic helimagnet will inevitably manifest itself when observing it by diffraction methods, for example, in neutron diffraction~\cite{Metlov2024}.

\section{Conclusions}
In this paper, we present a generalization of the model~\cite{M2025} of magnetic hopfions in a helimagnet, which allows for the azimuthal dependence of their profile function. It is shown that, despite the additional degrees of freedom, hopfions (of the first type~\cite{M2023_TwoTypes}) remain elliptically stable and retain an almost ideal spherical shape. The upper critical field of the lattice $h_{\uparrow}$ corresponding to its transition (of the second kind) to a state with uniform magnetization is calculated. Numerical calculations within the framework of the present extended model confirm the analytical expression~\eqref{eq:hup} for this critical field obtained earlier in~\cite{M2025}. The dependences of the hopfion dimensions (radius in the $\uZ=0$ plane and height along the $\uX=\uY=0$ line) as a function of external parameters -- normalized external field $\uh$ and anisotropy quality factor $\uq$ -- are calculated. It is shown that an increase in the external field approximately equally rescales all lattice periods and the size of individual hopfions, while uniaxial anisotropy makes the hopfions stretch along their axis ($\uX=\uY=0$) and leads to a consequent expansion of the equilibrium hopfion lattice.

The study was supported by the Russian Science Foundation grant No. 25-22-00076.

\section{APPENDIX: energy functions}
\label{sec:functions}
By substituting the trial function~\eqref{eq:model} into the expression for the free energy~\eqref{eq:energy}, integrating over the polar angle and renormalizing to the volume occupied by one hopfion in the lattice, the total normalized energy per unit volume $\uE\uC/\uD^2$ can be represented as an integral of~\eqref{eq:energyF}. Introducing abbreviated notations for the profile function and its derivatives $\ue=\ue(\rho,\ut)$, $\uer=\partial \ue(\rho,\ut)/\partial\rho$, $\uet=\partial \ue(\rho,\ut)/\partial\ut$, the energy functions $\upex$, $\updm$, $\upa$ and $\upz$ included in~\eqref{eq:energyF} can be represented as follows.

For the exchange energy:
\begin{equation*}
\upex=\frac{4\pi\sin\ut}{1+\cos\alpha}
\left[
\frac{b_{\mathrm{EX},0}}{(1-2p)^4}+
\frac{b_{\mathrm{EX},1}\uet}{(1-2p)^3}+
\frac{b_{\mathrm{EX},2}\uet^2}{(1-2p)^2}
\right],
\end{equation*}
where  $p=\ue(1-\ue)$, $\gamma=\tan(\alpha/2)$,
\begin{align*}
b_{\mathrm{EX},0} = & c_{\mathrm{EX},0}+c_{\mathrm{EX},1}\cos2\ut +c_{\mathrm{EX},2}\cos4\ut, \\
c_{\mathrm{EX},0} = & d_{\mathrm{EX},0} + d_{\mathrm{EX},1} \cos \alpha, \\ 
d_{\mathrm{EX},0} = & 2 (1-2 p)^2 \left(\rho^2 \uer^2+3 p^2\right), \\
d_{\mathrm{EX},1} = & (1\!-\!2 p)^2 \rho^2 \uer^2+2 (1-2 e (e (2 e\!-\!3)+2)) p \rho \uer+\\
&(4 p (2 p-5)+5) p^2, \\
c_{\mathrm{EX},1} = & d_{\mathrm{EX},2} + d_{\mathrm{EX},3} \cos \alpha, \\ 
d_{\mathrm{EX},2} = & 2 (1-2 p)^2 \left(p-\rho \uer\right) \left(\rho \uer+p\right),\\
d_{\mathrm{EX},3} = & 2 \left(-(1-2 p)^2 \rho^2 \uer^2+4 (3 p-1) p^3+p^2\right),\\
c_{\mathrm{EX},2} = &\left(2 p \rho \uer-2 e p-\rho \uer+p\right)^2 \cos\alpha, \\
b_{\mathrm{EX},1} = & c_{\mathrm{EX},3}\sin2\ut +c_{\mathrm{EX},4}\sin4\ut, \\
c_{\mathrm{EX},3} = & 4 \left((1-2 p) \rho \uer \cos \alpha -2 e p+p\right), \\
c_{\mathrm{EX},4} = & 2 \left(2 p \rho \uer-2 e p-\rho \uer+p\right) \cos\alpha, \\
b_{\mathrm{EX},2} = & -4 (2 p-1)(\cos \alpha \cos 2 \ut+1) \sin ^2\ut.
\end{align*}

For the Dzyaloshinskii-Moriya energy:
\begin{align*}
\updm= & 4\pi \rho \sin\ut
\left[
\frac{b_{\mathrm{EX},3}}{(1-2p)^4}+
\frac{b_{\mathrm{EX},4}\uet}{(1-2p)^3}
\right],\\
b_{\mathrm{EX},3} = & c_{\mathrm{EX},5}+c_{\mathrm{EX},6}\cos2\ut +c_{\mathrm{EX},7}\cos4\ut, \\
c_{\mathrm{EX},5} = & 
\left((2 p (\gamma  p\!+\!p\!-\!2)+1) \left(\rho \uer - p \left(2 \rho \uer\!+\!6 e\!-\!3\right)\right)\right),\\
c_{\mathrm{EX},6} = & p \left(2 e (4 p ((2 \gamma -3) p+1)-1)+\right.\\
&\left.4 p \left(p \left(-2 \gamma +2 \rho \uer+3\right)-3 \rho \uer-1\right)+\right. \\
&\left.6 \rho \uer+1\right)-\rho \uer \\
c_{\mathrm{EX},7} = & 2 (\gamma -1) p^2 \left(2 p \rho \uer-2 e p-\rho \uer+p\right),\\
b_{\mathrm{EX},3} = &
(2 p-1) (4 p (\gamma  p+1)-1) \sin 2\theta+ \\
&2 p^2 (\gamma -2 (\gamma -1) p-1) \sin 4 \theta.
\end{align*}
And finally, for the magnetic anisotropy energy and the Zeeman energy:
\begin{align*}
\upa= & 2 \pi  \rho^2 \sin \theta \left(1-\frac{(1-4 p (1-p \cos 2 \theta ))^2}{(1-2 p)^4}\right), \\
\upz= & \frac{16 \pi  p^2 \rho^2 \sin ^3\theta}{(1-2 p)^2},
\end{align*}
which do not depend on either $\gamma$ or the derivatives of the profile function.

If one sets $\uet=0$ in each of these functions and integrates over $\ut\in[0,\pi]$, the resulting expressions will exactly coincide with the corresponding functions for the simpler model~\cite{M2025}.

\bibliographystyle{jetp}

\end{document}


\def\TitleSkip{\vskip 5mm}
\def\AuthorsSkip{\bigskip}
\def\AffilSkip{\medskip}
\def\AffiliationsSkip{\medskip}
\def\DatesSkip{\bigskip}
\def\DOISkip{\vskip -5mm}
\def\LFVS{\vskip -5mm}
\def\ELFVS{\vspace*{0mm}}

\title{Управление решёткой хопфионов в гелимагнетике при помощи магнитного поля и анизотропии}

\rtitle{Влияние магнитного поля и анизотропии\dots}

\affiliation{Донецкий физико-технический институт им. А.А.~Галкина\\ 283048, г. Донецк, ДНР, Россия}

\author{К.~Л.}{Метлов}\email{metlov@donfti.ru}
\author{А.~С.}{Тарасенко}
\author{Ю.~А.}{Безус}
\author{М.~М.}{Гордей}

\dates{21 февраля 2025 г.}{*}
\abstract{Рассмотрено обобщение микромагнитной модели хопфионов в гелимагнетике с двумерной (допускающей как радиальную, так и азимутальную зависимость) функцией профиля. Расчёты подтверждают эллиптическую стабильность хопфионов и ранее полученное аналитическое выражение для верхнего критического поля их решётки. Показано, что в анизотропном гелимагнетике решётка хопфионов испытывает растяжение в направлении оси анизотропии, а её периоды зависят от величины приложенного внешнего магнитного поля и константы одноосной анизотропии материала.}

\DOI 10.7868/XXXXXXXXXXXXXXX

\maketitle
	
\section*{Введение}

Магнитные топологические солитоны~\cite{KIK83-RU} играют ключевую роль практически во всех приложениях магнетизма от сердечников трансформаторов, сенсоров, магнитных пластин в жестких дисках до твердотельной магнитной памяти и наномасштабных генераторов микроволн. С точки зрения топологии, все они соответствуют непрерывным отображениям сферы на сферу. При этом, целевая сфера --- всегда $S^2$, на которой лежат концы векторов локальной намагниченности фиксированной длины $|\uvM(\uvr)|=\uMs$. Исходная же сфера является отображением физического пространства $\uvr$ с наложенными периодическими граничными условиями в бесконечности. Известно, что в одном ($S^1\rightarrow S^2$) и в двух ($S^2\rightarrow S^2$, \cite{BP75-RU}) измерениях такие отображения распадаются на гомотопические классы, пронумерованные целыми числами. Так в векторном поле намагниченности возникают дискретные локализованные объекты (доменные границы~\cite{Hubert_book_walls-RU}, цилиндрические магнитные домены~\cite{BG88-RU}, скирмионы~\cite{BY1989-RU}, магнитные вихри~\cite{UP93,M10}) а номер класса играет роль количества этих топологических солитонов.

Долгое время в математике считалось, что все отображения $S^3\rightarrow S^2$ гомотопически эквивалентны (что означало бы невозможность существования трёхмерных топологических солитонов). Однако в 1931-м году Хайнц Хопф привел пример отображения~\cite{Hopf1931}, не сводящегося ко всем остальным. Этот пример был обобщён Уайтхедом~\cite{whitehead1947}, который показал, что и в этом случае отображения распадаются на пронумерованные целыми числами гомотопические классы, ввёл соответствующий индекс Хопфа $\uHopf$ (оригинальный контр-пример~\cite{Hopf1931} имеет $\uHopf=1$) и построил отображения с произвольными $\uHopf$. Идея поиска соответствующих этим отображениям солитонов в физике была высказана Людвигом Фаддеевым~\cite{Faddeev1976}, а конкретно в векторных полях локальной  намагниченности ферромагнетиков Дзялошинским и Ивановым~\cite{DI79-RU}. Но эта идея так и не реализовалась тогда в магнетизме в силу теоремы Хобарта-Деррика~\cite{Hobart1963,Derrick1964}, доказавших нестабильность трёхмерных солитонов в классическом ферромагнетике.

В последние годы наблюдается новый всплеск интереса к трёхмерным топологическим солитонам~\cite{RKBDMB2022} в хиральных магнетиках (выходящих за пределы применимости~\cite{BK2011} теоремы Хобарта-Деррика). Это связано с развитием методов наблюдения распределений намагниченности на нано-масштабах, таких как рентгеновская микроскопия высокого разрешения~\cite{KRRCVDCHSPFHSF2021}, векторная магнитная нано-томография~\cite{CMSGHBRHCG2020} и магнитное малоугловое рассеяние нейтронов~\cite{Metlov2024}, для которых магнитные хопфионы были бы идеальным объектом исследований. Стимулом также является потенциал применения хопфионов в трёхмерной спинтронике~\cite{BK2011,RKBDMB2022,Gubbiotti2024roadmap}.

Появились первые экспериментальные наблюдения хопфионов и хопфион-подобных состояний в ограниченной геометрии~\cite{KRRCVDCHSPFHSF2021,zheng2023,Yu2023}. Они также воспроизводятся качественно и в численных экспериментах~\cite{lake2018,TS2018,sutcliffe2018,WQB2019,RKBDMB2022}. Тем не менее, это не свободно стоящие объемные хопфионы, а их существование зависит от внешней стабилизации и удержания. Обычно это достигается путем помещения хопфион-содержащей пленки между двумя магнитными слоями с сильной перпендикулярной магнитной анизотропией~\cite{TS2018}. Но всё же истинный потенциал хопфионов может быть реализован только в полностью трехмерном «объемном» окружении.

Нашей отправной точкой здесь является вариационная модель объёмных хопфионов в гелимагнетике~\cite{M2023_TwoTypes} на базе стандартного микромагнитного гамильтониана с учётом взаимодействия Дзялошинского-Мория. Она предсказывает существование двух (топологически эквивалентных) типов магнитных хопфионов. Причём в~\cite{M2025}, один из них является стабильным, относительно эллиптической деформации, а второй -- нет. Хопфионы второго типа стремятся безгранично расшириться вдоль направления внешнего магнитного поля, сохраняя стабильный размер в перпендикулярном направлении. Это расширение можно остановить вводя в систему слой с повышенной одноосной магнитной анизотропией~\cite{TS2018}. Хопфионы же первого типа такой искусственной стабилизации не требуют.

В настоящей работе мы обобщили модель~\cite{M2025}, введя в пробную функцию дополнительное измерение (соответствующее дополнительному бесконечному набору вариационных параметров). Такая расширенная модель в пределе сводится к исходной, но, по определению (за счёт большего количества свободных параметров), даёт лучшее приближение к энергии точного решения. Проведена численная оценка верхнего критического поля хопфионной решётки, которое оказалось в полном (с учётом численной погрешности) согласии с полученным ранее аналитическим выражением~\cite{M2025}. Рассмотренная расширенная модель также позволяет ``отвязать'' внутреннюю структуру хопфиона от его внешнего контура, а значит, открывает возможность моделирования контактного (обменного, поскольку взаимодействие Дзялошинского-Мория тоже является его частью) расталкивания между хопфионами в решётке. Это позволило исследовать влияние внешнего поля и анизотропии на параметры (периоды) равновесной решётки хопфионов.

\section{Модель}	
\label{sec:model}
Рассмотрим классический гелимагнетик со свободной энергией на единицу объема $\uE=\lim_{V\rightarrow\infty} (1/V) \uiiint_V F \ud^3\uvr$ с плотностью
\begin{align}
        \uF = &\frac{\uC}{2}\sum_{i=\uX,\uY,\uZ} \left|\uGrad\umi\right|^2 +\uD\, \uvm\cdot[\uGrad\times\uvm] - \nonumber \\ 
        &\umuZ \uMs \left(\uvm \cdot \uvH\right) - \uK\left(\uvm \cdot \uvs\right)^2, 
        \label{eq:energy} 
\end{align}

где ${\bf m}({\bf r})={\bf M}({\bf r})/\uMs$, $|\uvm|=1$, $\uC=2A$ --- обменная жесткость в Дж/м, $\uD $ --- параметр взаимодействия Дзялошинского-Мория (ДМ)~\cite{BS1970,BJ1980} в $Дж/м^2$,\ $\uK$\ \textemdash \ константа\ одноосной\ (направленной вдоль $\uvs$) анизотропии  в $Дж/м^3 $, $\uvH$ --- внешнее магнитное поле в А/м, а $\umuZ=4\pi\times{10^{-7}}$ Гн/м --- проницаемость вакуума. Для простоты предположим, что поле и ось анизотропии параллельны, и выберем декартову систему координат так, что это направление совпадает с осью $\uO\uZ$: $\uvH=\{0,0,H\}$, $\uvs=\{0,0,1\}$. Задача теперь состоит в том, чтобы найти векторное поле $\uvm(\uvr)$, которое минимизирует $\uE$ и имеет $\uHopf>0$.

Точное решение данной задачи на сегодня не известно. Поэтому воспользуемся методом Ритца, выбрав в качестве отправной точки следующую пробную функцию для векторного поля намагниченности:
\begin{subequations}
\label{eq:model}
\begin{align}
\label{eq:stereogr}
&\{\umx+\imath\umy,\umz\}=\{2w, 1 - |w|^2\}/(1+|w|^2),\\
\label{eq:whitehead}
&w=\imath u/v,\\
\label{eq:E3S3}
 &u=\frac{2(\uXp+\imath \uYp)\uR}{\urp^2+\uR^2},
 \ 
 v=\frac{\uR^2-\urp^2+\imath 2 \uZp \uR}{\urp^2+\uR^2},
\\
\label{eq:physE3}
&\uvrp=\uR\,\frac{\uvr}{r}\frac{\ue}{1-\ue},\ \ue=\ue\left(\frac{r}{\uR},\arctan\frac{\uZ}{\sqrt{r^2-\uZ^2}}\right)
\end{align}
\end{subequations}
$\uvr=\{\uX,\uY,\gamma\uZ\}$, $r=|\uvr|$, $\uvrp=\{\uXp,\uYp,\uZp\}$, $\urp=|\uvrp|$. Она определяет $\uvm(\uvr)$ с индексом Хопфа $\uHopf=1$, параметризованную (пока неизвестной) функцией профиля $\ue(\rho,\theta)$, $\rho=\ur/\uR$, удовлетворяющей граничным условиям $\ue(0,\theta)=0$, $\ue(1,\theta)=1$, и двумя скалярными параметрами -- радиусом хопфиона $\uR$ и фактором формы (отношению размера в направлении перпендикулярном магнитному полю к размеру в направлении поля)  $\gamma$. Хопфион полностью содержится внутри эллипсоида $\rho<1$, а на его границе ($\rho=1$) имеет намагниченность $\uvm=\{0,0,1\}$, которая совпадает с намагниченностью снаружи хопфиона. Когда $|\gamma|<1$, эллипсоид вытянут вдоль оси $\uO\uZ$.

Ключевым в этой пробной функции является анзац Уайтхеда~\eqref{eq:whitehead}, который записан здесь в частном случае $\uHopf=1$. В общем случае~\cite{whitehead1947} этот анзац имеет вид $w=\imath u^m/v^n$, где $n$ и $m$ -- взаимно простые целые числа, что соответствует распределениям намагниченности с индексом Хопфа $\uHopf=m n$. Множитель $\imath$ в~\eqref{eq:whitehead} выбирает первый из двух типов хопфионов~\cite{M2023_TwoTypes} (хопфионы второго типа соответствуют множителю $-\imath$ и требуют дополнительной стабилизации, компенсирующей их эллиптическую нестабильность~\cite{M2025}). Хопфионы с $\uHopf=1$ являются аксиально симметричными относительно $\uO\uZ$ и состоят из круговых, обёрнутых вокруг $\uO\uZ$, вихревой и антивихревой нитей с противоположной намагниченностью в центрах их ядер. У рассмотренных здесь хопфионов первого типа внешняя антивихревая нить обернута вокруг внутренней вихревой как показано на
\begin{figure}[tb]\vspace{1mm}
\begin{center}
\hspace{0mm}
\includegraphics[width=\columnwidth]{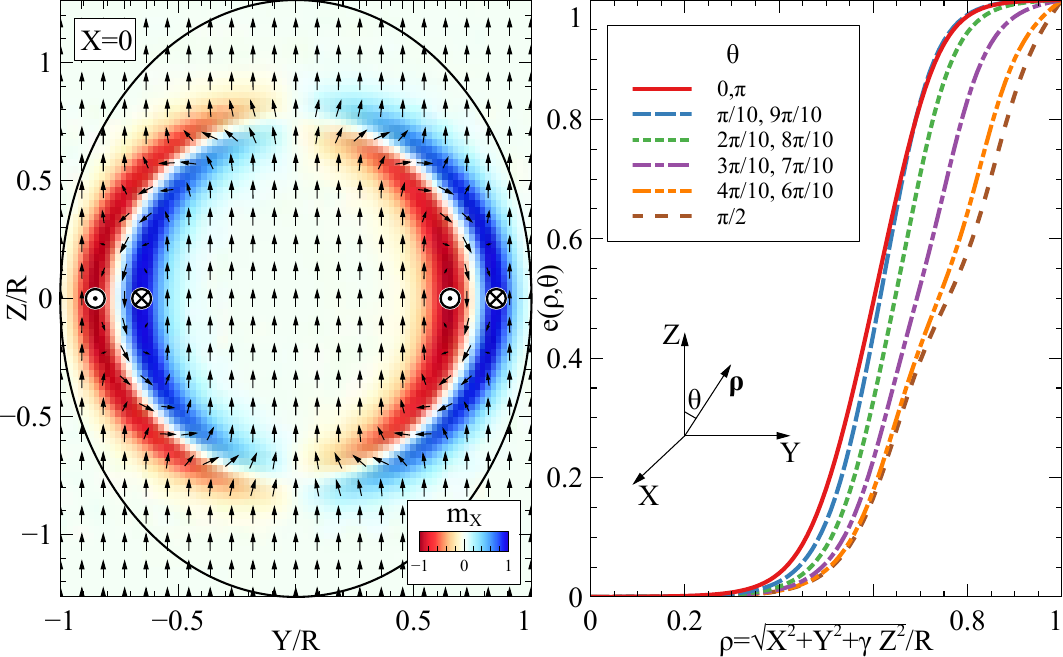}
\end{center}
\caption{\hspace{0mm}\footnotesize Сечение распределения намагниченности (слева) и некоторые срезы функции профиля $e(x,\theta)$ (справа) равновесного хопфиона с $\uHopf=1$ (его общая хиральность соответствует материалу с $\uD<0$) при $q=1$ и $h=0.1$}
	\label{fig:H1}
	\vspace{5mm}
\end{figure}	
Рис.~\ref{fig:H1} слева. У хопфионов второго типа порядок нитей меняется на противоположный. Анзац Уайтхеда~\eqref{eq:whitehead} определяет отображение сферы $S^3$, параметризованной двумя комплексными координатами $u$ и $v$ (такими что $|u|^2+|v|^2=1$), на сферу Римана $S^2$, параметризованную комплексной координатой $w$. Формулы~\eqref{eq:E3S3} выражают $u$ и $v$ в терминах координат в расширенном евклидовом пространстве $\{\uXp,\uYp,\uZp\}$, а стереографическая проекция~\eqref{eq:stereogr} связывает $w$ с компонентами вектора намагниченности фиксированной длины $|\uvm|=1$. Все вместе уравнения~\eqref{eq:stereogr},~\eqref{eq:whitehead},~\eqref{eq:E3S3} определяют векторное поле $\uvm(\uXp,\uYp,\uZp)$, содержащее один $\uHopf=1$ хопфион, заполняющий всё пространство $\uvrp$ целиком. Эти формулы универсальны в том смысле, что любые $\uHopf=1$ хопфионы (не только магнитные) обязательно сводятся к ним некоторым непрерывным отображением (гомотопией).

Основные приближения рассматриваемой модели содержатся в уравнении~\eqref{eq:physE3}, задающем отображение физического пространства $\{\uX,\uY,\uZ\}$ на расширенное евклидово пространство $\{\uXp,\uYp,\uZp\}$. Оно полностью локализует хопфион внутри эллипсоида $\rho<1$, отображая (за счёт граничного условия $\ue(1,\ut)=1$) точки на границе $\rho=1$ в бесконечно удалённую точку пространства $\uvrp$, где вектор намагниченности $\uvm=\{0,0,1\}$ направлен параллельно оси $\uO\uZ$. Появление $\ue$ в числителе~\eqref{eq:physE3} улучшает оригинальную вариационную модель~\cite{M2023_TwoTypes} и обсуждается в~\cite{M2025}. Ключевым отличием вариационной модели в данной работе является учёт возможной азимутальной зависимости профиля $\ue(\rho,\ut)$ хопфиона, в то время как в более ранних исследованиях профиль считался не зависящим от $\ut$.

Предположим теперь, что такие локализованные хопфионы выстроены в бесконечную трёхмерную плотноупакованную решётку (FCC или HCP) с объёмом $V=4\sqrt{2}R^3\gamma$, приходящимся на один хопфион. При том, что каждый отдельный хопфион имеет $\uHopf=1$, индекс Хопфа решётки в целом равен бесконечности. Полную, нормированную на $D^2/C$, энергию единицы её объёма можно представить в виде $\uE\uC/\uD^2=\int_0^1\int_{0}^{\pi}{\cal F}\ud\ut\ud \rho$, где
\begin{equation}
 \label{eq:energyF}
 {\cal F}= \frac{1}{\kappa} \left[\nu^2\upex+\nu\,\updm+\frac{q}{2}(\upa\!-\!\kappa)+h(\upz\!-\!\kappa)\right],
\end{equation}
$\kappa=4\sqrt{2}$, $\nu=\uC/(\uD \uR)$ --- безразмерный параметр, обратно пропорциональный размеру хопфиона $\uR$, $\uh=\umuZ\uMs\uC\uH/\uD^2$ --- нормированное внешнее поле, $\uq=2\uC\uK/\uD^2$ --- нормированный фактор качества одноосной анизотропии. Выражения для функций $\upex$, $\updm$, $\upa$ и $\upz$ даны в дополнительном разделе~\ref{sec:functions}. Энергия $\uE$ является функционалом от профиля хопфиона $\ue(\rho,\ut)$ и зависит от двух дополнительных безразмерных скалярных параметров --- $\nu$ и $\gamma$.

Минимизация энергии $\uE$ в данной работе проводилась численно в два этапа. На первом этапе, методом конечных элементов  рассчитывался равновесный профиль хопфиона  $\ue(\rho,\ut)$ при заданных $\uq$, $\uh$, $\nu$ и $\gamma$ как решение уравнения Эйлера-Лагранжа:
\begin{equation}
 \label{eq:EulerLagrange}
 \frac{\partial \uFc}{\partial \ue} -
 \frac{\partial}{\partial\rho}
 \frac{\partial \uFc}{\partial\uer} -
 \frac{\partial}{\partial\theta}
 \frac{\partial \uFc}{\partial\uet} = 0,
 \end{equation}
где $\uer=\partial \ue(\rho,\ut)/\partial\rho$, $\uet=\partial \ue(\rho,\ut)/\partial\ut$ с граничными условиями $\ue(0,\ut)=0$, $\ue(1,\ut)=1$. Для этого использовалась функция {\tt NDSolve`FEM`} системы компьютерной алгебры Mathematica\texttrademark{}. Для стабилизации численного счёта в уравнение Эйлера-Лагранжа вводился дополнительный диффузионный член $0.0025(\partial^2\ue/\partial\rho^2+\partial^2\ue/\partial\theta^2)$, который практически не влияет на значение энергии хопфиона, но хорошо подавляет высокочастотные численные флуктуации профиля около $e\approx0$. Затем, для получившегося профиля вычислялась нормированная энергия $\uE\uC/\uD^2$  посредством прямого численного интегрирования~\eqref{eq:energyF} и минимизировалась численно при помощи функции {\tt FindMinimum} по оставшимся двум скалярным параметрам $\nu$ и $\gamma$. Пример рассчитанного таким образом хопфиона и соответствующие функции профиля (для некоторых значений $\ut$) приведены на Рис.~\ref{fig:H1}.

\section{Результаты}

Уже из Рис.~\ref{fig:H1} видна основная особенность рассмотренной модели, заключающаяся в том, что ядро хопфиона (область сильно неоднородной намагниченности) больше не привязано к его внешнему контуру $\rho=1$ (показанному черным эллипсом на Рис.~\ref{fig:H1} слева). Над и под ядром формируются области почти однородной намагниченности, но таким образом, что само ядро хопфиона (и этот эффект сохраняется для других $\uq$ и $\uh$) сохраняет практически идеальную сферическую форму (притом, что внешний контур хопфиона приобретает значительную эллиптическую деформацию). Это находится в согласии с выводами, полученными в рамках более простой модели~\cite{M2025}, и подтверждает вывод об эллиптической стабильности хопфионов.

В области $\uq>0$, $\uh>0$, когда хопфионы наиболее энергетически выгодны, их область стабильности ограничена верхним критическим полем (обращающимся в $0$ при некотором предельном значении фактора качества анизотропии $\uq$). В слабых (подкритических) внешних полях энергия равновесной решётки хопфионов ниже энергии состояния с однородной намагниченностью. На критической линии эти энергии становятся равны и происходит взаимное (безгистерезисное) превращение решётки хопфионов в состояние однородной намагниченности посредством фазового перехода второго рода.

При приближении к критической линии энергетический минимум для радиуса хопфиона превращается в седловую точку и исчезает, а сам радиус стремится к бесконечности (однородно намагниченное ядро хопфиона при этом расширяется и заполняет всё пространство). Это значит, что численная процедура поиска минимума полной энергии хопфиона по $\nu$ и $\gamma$ перестаёт работать вблизи критической линии. Тем не менее, к ней можно подобраться достаточно близко, чтобы использовать линейную (по полю $\uh$) экстраполяцию энергии до точки, где энергия хопфионов становится в точности равной энергии состояния однородной намагниченности. Результаты этой экстраполяции показаны точками на
\begin{figure}[tb]\vspace{1mm}
\begin{center}
\hspace{0mm}
\includegraphics[width=\columnwidth]{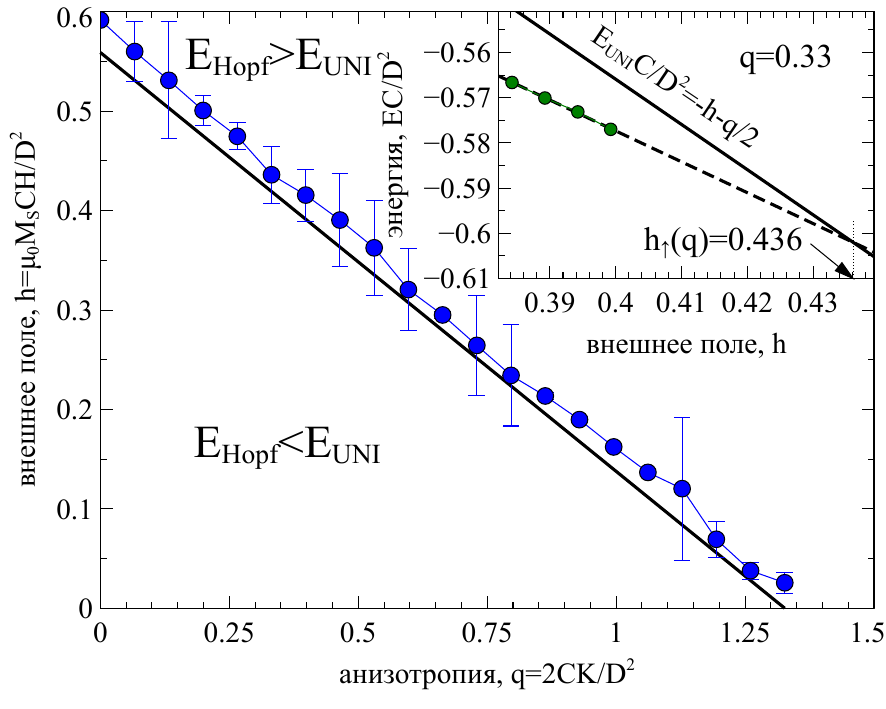}
\end{center}
\caption{\hspace{0mm}\footnotesize Верхнее критическое поле решетки хопфионов, соответствующее её переходу в состояние с однородной намагниченностью; вставка иллюстрирует процедуру экстраполяции на примере одной точки с $q=0.33$}
	\label{fig:hcrupper}
	\vspace{5mm}
\end{figure}	
Рис.~\ref{fig:hcrupper}. Изображенные погрешности учитывают лишь статистический разброс использованных при экстраполяции при каждом значении $\uq$ четырёх точек.
Сплошной линией на Рис.~\ref{fig:hcrupper} показано критическое поле, вычисленное по формуле
\begin{equation}
\label{eq:hup}
 h_{\uparrow}=\frac{4 (2+\pi ) (3 \pi -4) q-15 \pi ^2}{60 (\pi -4) (2+\pi )},
\end{equation}
полученной в~\cite{M2025}. Заметное систематическое превышение численной оценки критического поля, по-видимому, связано с дополнительно введённым в уравнение диффузионным членом. Но в целом, можно сказать, что численная оценка критического поля находится в очень хорошем согласии с аналитической формулой.   
Учитывая, что такое согласие присутствует для трёх разных моделей хопфионов:~\cite{M2023_TwoTypes}, \cite{M2025} и нынешней -- можно сказать, что оценка~\eqref{eq:hup}, если и не является точным результатом, то очень близка к нему.

Описанная в разделе~\ref{sec:model} численная процедура позволяет вычислить равновесные значения параметров $\nu$ и $\gamma$ во всей области стабильности хопфионов, показанной на Рис.~\ref{fig:hcrupper}. Из них можно оценить влияние внешнего поля и анизотропии на радиус (в плоскости $\uZ=0$) $\propto1/\nu$  и высоту (размер в направлении поля) $\propto1/(\gamma\nu)$ хопфиона. Эти зависимости изображены на
\begin{figure}[tb]\vspace{1mm}
\begin{center}
\hspace{0mm}
\includegraphics[width=\columnwidth]{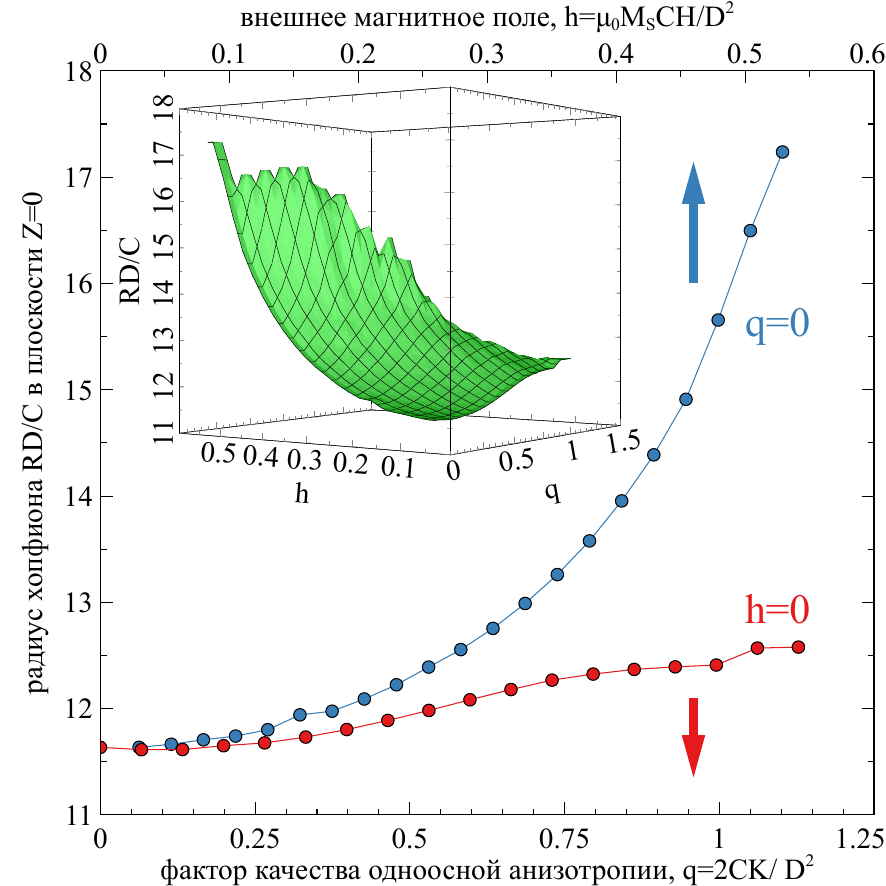}
\end{center}
\caption{\hspace{0mm}\footnotesize Радиус хопфиона как функция приложенного магнитного поля и фактора качества одноосной анизотропии при $q=0$ и $h=0$; на вставке показан полный график во всей области стабильности хопфионов при $h>0$, $q>0$}
	\label{fig:width}
	\vspace{2mm}
\end{figure}	
\begin{figure}[tb]\vspace{0mm}
\begin{center}
\hspace{0mm}
\includegraphics[width=\columnwidth]{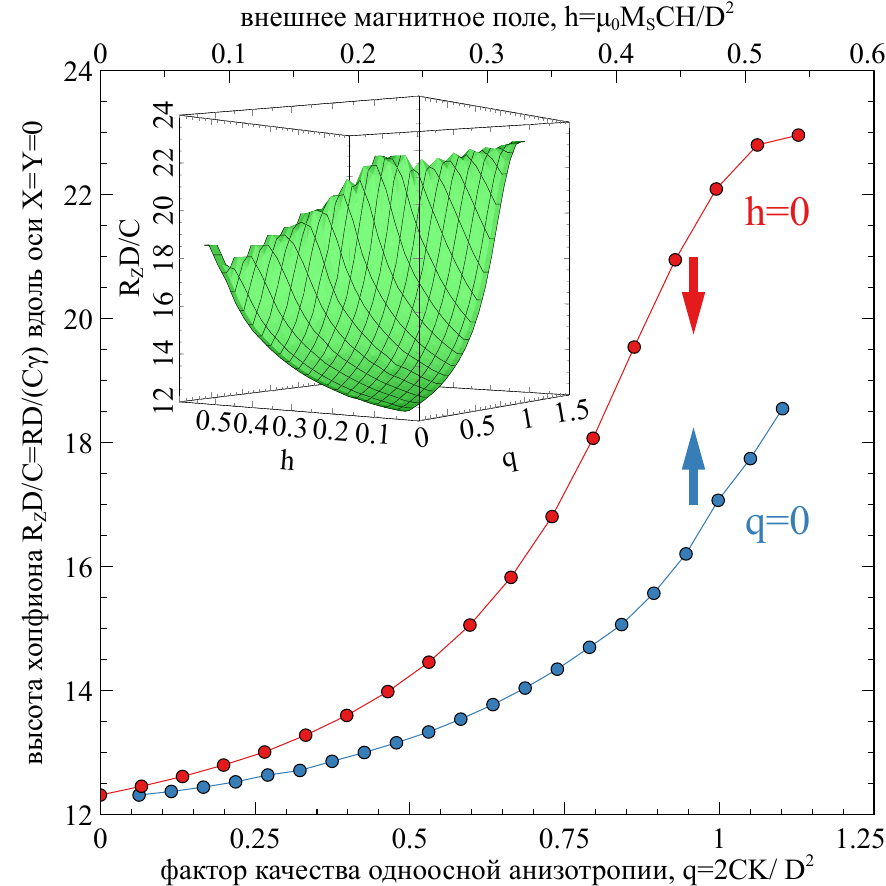}
\end{center}
\caption{\hspace{0mm}\footnotesize Высота хопфиона как функция приложенного магнитного поля и фактора качества одноосной анизотропии; композиция графика идентична Рис.~\ref{fig:width}}
	\label{fig:height}
	\vspace{4mm}
\end{figure}	
Рис.~\ref{fig:width} и Рис.~\ref{fig:height}. Видно (особенно хорошо на графиках при $\uq=0$), что изменение внешнего поля вызывает примерно одинаковое изменение радиуса и высоты хопфиона, а изменение фактора качества анизотропии на радиусе хопфиона практически не отражается (см. Рис.~\ref{fig:width} при $\uh=0$), но сильно влияет на его высоту (см. Рис.~\ref{fig:height} при $\uh=0$). Сам хопфион (та его область, где намагниченность существенно отклоняется от однородной) сохраняет почти идеальную сферическую форму, меняется лишь его внешний контур (показан эллипсом на Рис.~\ref{fig:H1}). Тем не менее, даже малые отклонения намагниченности внутри контура хопфиона от идеальной вертикальной ($\uvm=\{0,0,1\}$ за пределами хопфиона) производят обменные силы, препятствующие сближению хопфионов. 

Поскольку энергия стабильного хопфиона ниже энергии однородно намагниченного состояния, решётка хопфионов будет плотно упакованной, чтобы минимизировать пространство (с однородной намагниченностью) между ними. Задача о плотной упаковке эллипсоидов имеет ряд нетривиальных решений~\cite{Donev2004}, не сводящихся к перемасштабированной решётке плотноупакованных сфер. Однако, во всех этих необычных случаях эллипсоиды не являются соосными. Хопфионы же в отсутствии анизотропии имеют почти идеальную сферическую форму, и приобретают заметное эллиптическое искажение только в анизотропном магнетике. Но эта же анизотропия навязывает им выделенное направление, а значит мешает реализации более сложных плотноупакованных состояний~\cite{Donev2004}.

В целом, это означает, что внешнее поле однородно масштабирует решётку плотноупакованных сфер (и размеры каждого хопфиона в отдельности), а наличие одноосной анизотропии деформирует её сильнее в одном направлении. Такое растяжение решётки хопфионов в анизотропном гелимагнетике неизбежно проявится при наблюдении её дифракционными методами, например в нейтронографии~\cite{Metlov2024}.

\section{Выводы}

В работе представлено обобщение модели~\cite{M2025} магнитных хопфионов в гелимагнетике, допускающее азимутальную зависимость их функции профиля. Показано, что, несмотря на дополнительные степени свободы, хопфионы (первого типа~\cite{M2023_TwoTypes}) остаются эллиптически стабильными и сохраняют практически идеальную сферическую форму. Рассчитано верхнее критическое поле решетки $h_{\uparrow}$, соответствующее её переходу (II-го рода) в состояние с однородной намагниченностью. Численные расчёты в рамках настоящей расширенной модели подтверждают ранее полученное в~\cite{M2025} аналитическое выражение~\eqref{eq:hup} для этого критического поля. Рассчитаны зависимости размеров (радиуса в плоскости $\uZ=0$ и высоты вдоль линии $\uX=\uY=0$) хопфиона как функции внешних параметров -- нормированных внешнего поля $\uh$ и фактора качества анизотропии $\uq$. Показано, что увеличение внешнего поля примерно одинаково перемасштабирует все периоды решётки и размер отдельных хопфионов, в то время как одноосная анизотропия заставляет хопфионы вытягиваться вдоль своей оси ($\uX=\uY=0$) и приводит к соответствующему растяжению равновесной решётки хопфионов.

Исследование выполнено за счет гранта Российского научного фонда № 25-22-00076.

\section{ПРИЛОЖЕНИЕ: энергетические функции}
\label{sec:functions}
Подстановкой пробной функции~\eqref{eq:model} в выражение для свободной энергии~\eqref{eq:energy} интегрированием по полярному углу и перенормировкой на объем, занимаемый одним хопфионом в решётке, полную нормированную энергию на единицу объёма $\uE\uC/\uD^2$ можно представить в виде интеграла от~\eqref{eq:energyF}. Вводя сокращённые обозначения для функции профиля и её производных $\ue=\ue(\rho,\ut)$, $\uer=\partial \ue(\rho,\ut)/\partial\rho$, $\uet=\partial \ue(\rho,\ut)/\partial\ut$, входящие в~\eqref{eq:energyF} энергетические функции $\upex$, $\updm$, $\upa$ и $\upz$ можно представить следующим образом.

Для обменной энергии:
\begin{equation*}
\upex=\frac{4\pi\sin\ut}{1+\cos\alpha}
\left[
\frac{b_{\mathrm{EX},0}}{(1-2p)^4}+
\frac{b_{\mathrm{EX},1}\uet}{(1-2p)^3}+
\frac{b_{\mathrm{EX},2}\uet^2}{(1-2p)^2}
\right],
\end{equation*}
где $p=\ue(1-\ue)$, $\gamma=\tan(\alpha/2)$,
\begin{align*}
b_{\mathrm{EX},0} = & c_{\mathrm{EX},0}+c_{\mathrm{EX},1}\cos2\ut +c_{\mathrm{EX},2}\cos4\ut, \\
c_{\mathrm{EX},0} = & d_{\mathrm{EX},0} + d_{\mathrm{EX},1} \cos \alpha, \\ 
d_{\mathrm{EX},0} = & 2 (1-2 p)^2 \left(\rho^2 \uer^2+3 p^2\right), \\
d_{\mathrm{EX},1} = & (1\!-\!2 p)^2 \rho^2 \uer^2+2 (1-2 e (e (2 e\!-\!3)+2)) p \rho \uer+\\
&(4 p (2 p-5)+5) p^2, \\
c_{\mathrm{EX},1} = & d_{\mathrm{EX},2} + d_{\mathrm{EX},3} \cos \alpha, \\ 
d_{\mathrm{EX},2} = & 2 (1-2 p)^2 \left(p-\rho \uer\right) \left(\rho \uer+p\right),\\
d_{\mathrm{EX},3} = & 2 \left(-(1-2 p)^2 \rho^2 \uer^2+4 (3 p-1) p^3+p^2\right),\\
c_{\mathrm{EX},2} = &\left(2 p \rho \uer-2 e p-\rho \uer+p\right)^2 \cos\alpha, \\
b_{\mathrm{EX},1} = & c_{\mathrm{EX},3}\sin2\ut +c_{\mathrm{EX},4}\sin4\ut, \\
c_{\mathrm{EX},3} = & 4 \left((1-2 p) \rho \uer \cos \alpha -2 e p+p\right), \\
c_{\mathrm{EX},4} = & 2 \left(2 p \rho \uer-2 e p-\rho \uer+p\right) \cos\alpha, \\
b_{\mathrm{EX},2} = & -4 (2 p-1)(\cos \alpha \cos 2 \ut+1) \sin ^2\ut.
\end{align*}

Для взаимодействия Дзялошинского-Мория:
\begin{align*}
\updm= & 4\pi \rho \sin\ut
\left[
\frac{b_{\mathrm{EX},3}}{(1-2p)^4}+
\frac{b_{\mathrm{EX},4}\uet}{(1-2p)^3}
\right],\\
b_{\mathrm{EX},3} = & c_{\mathrm{EX},5}+c_{\mathrm{EX},6}\cos2\ut +c_{\mathrm{EX},7}\cos4\ut, \\
c_{\mathrm{EX},5} = & 
\left((2 p (\gamma  p\!+\!p\!-\!2)+1) \left(\rho \uer - p \left(2 \rho \uer\!+\!6 e\!-\!3\right)\right)\right),\\
c_{\mathrm{EX},6} = & p \left(2 e (4 p ((2 \gamma -3) p+1)-1)+\right.\\
&\left.4 p \left(p \left(-2 \gamma +2 \rho \uer+3\right)-3 \rho \uer-1\right)+\right. \\
&\left.6 \rho \uer+1\right)-\rho \uer \\
c_{\mathrm{EX},7} = & 2 (\gamma -1) p^2 \left(2 p \rho \uer-2 e p-\rho \uer+p\right),\\
b_{\mathrm{EX},3} = &
(2 p-1) (4 p (\gamma  p+1)-1) \sin 2\theta+ \\
&2 p^2 (\gamma -2 (\gamma -1) p-1) \sin 4 \theta.
\end{align*}
И, наконец, для энергии магнитной анизотропии и зеемановской энергии:
\begin{align*}
\upa= & 2 \pi  \rho^2 \sin \theta \left(1-\frac{(1-4 p (1-p \cos 2 \theta ))^2}{(1-2 p)^4}\right), \\
\upz= & \frac{16 \pi  p^2 \rho^2 \sin ^3\theta}{(1-2 p)^2},
\end{align*}
которые не зависят ни от $\gamma$, ни от производных функции профиля.

Если в каждой из этих функций положить $\uet=0$ и проинтегрировать по $\ut\in[0,\pi]$, получившиеся выражения в точности совпадут с соответствующими функциями для более простой модели~\cite{M2025}.

\bibliographystyle{jetp}